\documentclass[12pt]{iopart}
\usepackage[dvips,dvipdfm]{graphicx}
\usepackage{amssymb}
\usepackage{fancybox}
\usepackage{bm}
\renewcommand{\vec}[1]{\boldsymbol{#1}}
\begin{document}

\title[Typical performance of irregular LDGM codes for lossy compression]
{Typical performance of irregular low-density generator-matrix codes for lossy compression}
\author{Kazushi Mimura}
\address{
Faculty of Information Sciences, Hiroshima City University, 
3-4-1 Ohtsuka-Higashi, Asaminami-Ku, Hiroshima, 731-3194, Japan}
\ead{mimura@hiroshima-cu.ac.jp}

\begin{abstract}
We evaluate typical performance of irregular low-density generator-matrix (LDGM) codes, 
which is defined by sparse matrices with arbitrary irregular bit degree distribution and arbitrary check degree distribution, 
for lossy compression. 
We apply the replica method under one-step replica symmetry breaking (1RSB) ansatz to this problem. 
%\keywords{LDGM codes, typical performance, degree distribution, replica method}
\end{abstract}
\maketitle
% 89.90.+n  Other topics in areas of applied and interdisciplinary physics
% 02.50.-r  Probability theory, stochastic processes, and statistics
% 05.50.+q  Lattice theory and statistics (Ising, Potts, etc.) 
% 75.10.Hk  Classical spin models 

\section{Introduction}
%~~~~~~~~~~~~~~~~~~~~~~~~~~~~~~~~~~~~~~~~~~~~~~~~~~~~~~~~~~~~~~~~~~~~~
The channel coding can be considered as the dual problem of lossy source coding in rate-distortion theory \cite{Cover2006, Csiszar1981}. 
Matsunaga and Yamamoto showed that it is possible to approach the binary rate-distortion bound using LDPC codes \cite{Matsunaga2003}. 
In recent years, lossy source coding problem based on low-density generator-matrix (LDGM) codes is widely investigated. 
\par
This scheme can attain high performance very close to the Shannon bound, 
however it needs solinving a combinatorial optimization problem to obtain optimal source coding. 
Some practical encoding algorithms are proposed for this scheme, e.g., 
a belief-propagation-based encoder proposed by Murayama \cite{Murayama2004} and 
a survey-propagation-based encoder proposed by Wainwright and Maneva \cite{Wainwright2005}. 
\par
Performance of this scheme is also explored by various approaches. 
Murayama and Okada applied replica methods to evaluate performance of LDGM codes defined by regular sparse matrices for lossy compression \cite{Murayama2003}. 
Ciliberti et al. have used the cavity method to evaluate check-regular LDGM performance \cite{Ciliberti2005, Ciliberti2006}. 
On the other hand, Martinian and Wainwright derived rigorous upper bounds 
on the effective rate-distortion function of LDGM codes for the binary symmetric source \cite{Martinian2006}. 
Dimakis et al. derived lower bounds for check-regular LDGM codes \cite{Dimakis2007, Wainwright2007}. 
\par
With respect to irregular LDGM codes analyzed so far, 
elements of a reproduced message are given by exactly $K$ elements chosen at random from a codeword. 
This implies that previous analyses treat only the case where a bit degree distribution is Poissonian. 
An irregular bit and check degree distributions of a generator matrix are not optimized for lossy source coding. 
The goal of this paper is to evaluate how typical performance of irregular LDGM codes for lossy compression 
depends on a bit degree distribution and a check degree distribution.

\section{Background}
%~~~~~~~~~~~~~~~~~~~~~~~~~~~~~~~~~~~~~~~~~~~~~~~~~~~~~~~~~~~~~~~~~~~~~
\par
Let us first provide the concepts of the rate-distortion theory \cite{Cover2006}. 
Let $x$ be a binary i.i.d. source discrete which takes in a source alphabet $\mathcal{X}=\{ 0,1 \}$ with $\mathbb{P} [ x=0 ] = \mathbb{P} [ x=1 ] = 1/2$, 
where $\mathbb{P}$ represents the probability of its argument. 
An source message of $M$ random variables, $\vec{x}={}^t(x_1,\cdots,x_M) \in \mathcal{X}^M$, 
is compressed into a shorter expression, where the operator ${}^t$ denotes the transpose. 
The encoder describes the source sequence $\vec{x} \in \mathcal{X}^M$ by a codeword $\vec{z} = \mathcal{F}(\vec{x}) \in \mathcal{X}^N$. 
The decoder represents $\vec{x}$ by a reproduced message $\hat{\vec{x}} = \mathcal{G}(\vec{z}) \in \mathcal{X}^M$. 
Note that $M$ represents the length of a source sequence, while $N(<M)$ represents the length of a codeword. 
The code rate is $R=N/M$. 
The distortion between single letters is measured by the Hamming distortion defined by 
\begin{equation}
  d(x,\hat{x}) = \biggl\{
  \begin{array}{ll}
    0, & \; {\rm if} \; x = \hat{x}, \\
    1, & \; {\rm if} \; x \ne \hat{x},
  \end{array}
\end{equation}
and the distortion between $M$-bit sequences $\vec{x} \in \mathcal{X}^M$ and $\hat{\vec{x}} \in \mathcal{X}^M$ is measured 
by the averaged single-letter distortion as $d(\vec{x},\hat{\vec{x}})=\frac 1M \sum_{\mu=1}^M d(x_\mu,\hat{x}_\mu)$.
This results in the probability of error distortion, since  $\mathbb{E} [d(x,\hat{x})] = \mathbb{P} [x \ne \hat{x}]$, where $\mathbb{E}$ represents the expectation. 
The distortion associated with the code is defined as $D=\mathbb{E}[d(\vec{x},\hat{\vec{x}})]$, 
where the expectation is over the probability distribution on $\mathcal{X}^M \times \mathcal{X}^M$. 
A rate distortion pair $(R,D)$ is said to be {\it achievable} 
if there exists a sequence of rate distortion codes $({\cal F},{\cal G})$ with $\mathbb{E}[d(\vec{x},\hat{\vec{x}})] \le D$ in the limit $M\to\infty$. 
The rate distortion function $R(D)$is the infimum of rates $R$ such that $(R,D)$ is in the rate distortion region of the source for a given distortion $D$. 
The rate-distortion function of a Bernoulli($1/2$) i.i.d. source is given by 
\begin{equation}
  R(D)=1-h_2(D), 
  \label{eq:RDF}
\end{equation}
where $h_2(x)=-x\log_2(x)-(1-x)\log_2(1-x)$ is the binary entropy function.

\section{Lossy Compression Scheme}
%~~~~~~~~~~~~~~~~~~~~~~~~~~~~~~~~~~~~~~~~~~~~~~~~~~~~~~~~~~~~~~~~~~~~~
\par
An source message of $M$ random variables, $\vec{x} \in \mathcal{X}^M$, 
is compressed into a shorter expression, where the operator ${}^t$ denotes the transpose. 
The encoder describes the source sequence $\vec{x} \in \mathcal{X}^M$ by a codeword $\vec{z} = \mathcal{F}(\vec{x}) \in \mathcal{X}^N$. 
The decoder represents $\vec{x}$ by a reproduced message $\hat{\vec{x}} = \mathcal{G}(\vec{z}) \in \mathcal{X}^M$. 
The code rate is $R=N/M \le 1$. 
\par
Using a given $M \times N$ sparse matrix ${\cal A}=(a_{\mu i}) \in \{0,1\}^{M \times N}$, the decoder is defined as 
\begin{equation}
  \mathcal{G}(\vec{z}) = \mathcal{A} \vec{z} \quad ({\rm mod} \; 2). 
  \label{eq:decoder}
\end{equation}
The encoding is represented by 
\begin{equation}
  \mathcal{F}(\vec{x}) = \mathop{\rm argmin}_{\hat{\vec{z}}\in\mathcal{X}^N} d(\vec{x}, \mathcal{G}(\hat{\vec{z}})), 
\end{equation}
where $d$ is the distortion measure. 
In this paper, we use the Hamming distortion. 
Although the definition means that a computational cost of the encoding is of $O(e^N)$, 
we can utilize some suboptimal algorithms based on message passing to encode \cite{Murayama2004, Wainwright2005}.

\section{Analysis}
%~~~~~~~~~~~~~~~~~~~~~~~~~~~~~~~~~~~~~~~~~~~~~~~~~~~~~~~~~~~~~~~~~~~~~
\par
To simplify the calculations, we first introduce a simple isomorphism between the additive Boolean group $(\{0,1\},\oplus)$ and the multiplicative Ising group $(\{+1,-1\},\times)$ 
defined by $J \times \hat{J} = (-1)^{x \oplus \hat{x}}$, where $J,\hat{J} \in \{+1,-1\} = \mathcal{J}$ and $x,\hat{x} \in \{0,1\} = \mathcal{X}$. 
Hereafter, we use the following Ising (bipolar) representations : 
the Ising source message $\vec{J} \in \mathcal{J}^M$, 
the Ising reproduced message $\hat{\vec{J}} \in \mathcal{J}^M$ and 
the Ising codeword $\vec{\xi} \in \mathcal{J}^N$. 
The source bit can be described as a random variable with the probability: 
\begin{equation}
P_J(J) = \frac 12 \delta (J-1) + \frac 12 \delta(J+1), 
\end{equation}
where $\delta(x)$ denotes Dirac's delta function. 
The $\mu$-th element of the Ising reproduced message $\hat{J}_\mu$ is given by products of the elements of the tentative Ising codeword $\vec{s}\in\mathcal{J}^N$: 
\begin{equation}
  \hat{J}_\mu = \prod_{i \in \mathcal{L}(\mu)} s_i, 
\end{equation}
where $\mathcal{L}(\mu) = \{ i | a_{\mu i}=1 , \mathcal{A}=(a_{\mu i}) \}$. 
\par
The matrix $\mathcal{A}$ has $K_\mu$ nonzero elements in the $\mu$-th row and $C_i$ nonzero elements in the $i$-th column. 
We consider the source length and the codeword length to be infinite, while code rate $R$ is kept finite. 
The parameter $K_1 \cdots K_M$ and $C_1,\cdots,C_N$ are usually of $O(N^0)$, therefore the matrix $\mathcal{A}$ becomes very sparse. 
In densely constructed cases, we also assume that these parameters are not of $O(N^0)$ but $K, C_1 ,\cdots ,C_N \ll N$ holds. 
Counting the number of nonzero elements in the matrices leads to $K_1 + \cdots + K_M = C_1+\cdots+C_N$. 
The code rate is therefore $R = \tilde{K} / \tilde{C}$,
where $\tilde{K}=\frac 1N \sum_{\mu=1}^M K_\mu$ and $\tilde{C}=\frac 1N \sum_{i=1}^N C_i$. 
Code constructions are described by the connectivity parameter $\mathcal{D}^\mu_{i_1, \cdots, i_{K_\mu}} \in \{0,1\}$ 
which specifies a set of indices $i_1, \cdots, i_{K_\mu}$ corresponding to nonzero elements in the $\mu$-th row of the sparse matrix $\mathcal{A}$. 
The connectivity parameter is defined by 
\begin{equation}
  \mathcal{D}^\mu_{i_1, \cdots, i_{K_\mu}} =
  \left\{
  \begin{array}{ll}
    1, & \; {\rm if} \; \{i_1, \cdots, i_{K_\mu} \} = \mathcal{L}(\mu) \\
    0, & \; {\rm otherwise}
  \end{array}
  \right. .
\end{equation}
\par
An ensemble of codes is generated as follows. 
(i) Sets of $\{K_1,\cdots K_M\}$ and  $\{C_1,\cdots C_N\}$ are sampled independently from an identical distributions $P_K(K)$ and $P_C(C)$, respectively. 
(ii) The connectivity parameters $\mathcal{D}^\mu_{i_1, \cdots, i_{K_\mu}}$ are generated such that 
\begin{equation}
  \sum_{\mu=1}^M \sum_{\langle i_1=i, i_2, \cdots , i_{K_\mu} \rangle} \!\!\! \mathcal{D}^\mu_{i, i_2, \cdots, i_{K_\mu}} = C_i, 
  \label{eq:column_constraint}
\end{equation}
where $\sum_{\langle i_1=i, i_2, \cdots , i_{K_\mu} \rangle}$ denote 
the summation over $\{(i_2,\cdots,i_{K_\mu})$$\in\{1,\cdots,N\}^{{K_\mu}-1}$$|i_2<\cdots<i_{K_\mu},$$i_2 \!\ne\! i, \cdots, i_{K_\mu} \!\ne\! i\}$. 
\par
To analyze typical performance of rate-compatible LDGM codes for lossy compression, 
we apply a analytical method similar to references \cite{Murayama2003, Kabashima1999, Vicente2000, Nakamura2003, Mimura2006}. 
The Hamming distortion $d(\vec{J},\hat{\vec{J}})$ becomes 
$d(\vec{J},\hat{\vec{J}}) = \frac 12 -  \frac 1{2M} \sum_{\mu=1}^M J_\mu \{ \prod_{i \in \mathcal{L}(\mu)} s_i \}$, 
since $\vec{J}, \hat{\vec{J}} \in \mathcal{J}^M$. 
Using the connectivity parameter $\mathcal{D}^\mu_{i_1,\cdots,i_{K_\mu}}$, we can rewrite this Hamming distortion in the form: 
\begin{equation}
  d(\vec{J},\hat{\vec{J}}) = \frac 12 -  \frac 1{2M} \sum_{\mu=1}^M \sum_{\langle i_1,\cdots,i_{K_\mu} \rangle} \!\!\!\!\!\! \mathcal{D}^\mu_{i_1,\cdots,i_{K_\mu}} J_\mu s_{i_1} \cdots s_{i_{K_\mu}}, 
\end{equation}
where $\sum_{\langle i_1, \cdots , i_{K_\mu} \rangle}$ denote the summation over $\{(i_1,\cdots,i_{K_\mu})$$\in\{1,\cdots,N\}^{K_\mu}$$|i_1<\cdots<i_{K_\mu}\}$. 
We here define the Hamiltonian 
\begin{equation}
  \mathcal{H}(\vec{s},\vec{J}) = M d(\vec{J},\hat{\vec{J}}(\vec{s})), 
\end{equation}
to explore typical performance. 
The free energy is calculated from the partition function $Z(\beta)=\sum_{\vec{s}\in\mathcal{J}^N} \exp [ -\beta \mathcal{H}(\vec{s},\vec{J}) ]$. 
From the free energy, we can obtain a distortion between an original message and a reproduction message $D$ for a fixed code rate $R$. 
We follow the calculation of references \cite{Murayama2003, Kabashima1999, Krauth1989, Tanaka2003, Yano2007, Yano2008}.

\subsection{Replica symmetric solution}
%~~~~~~~~~~~~~~~~~~~~~~~~~~~~~~~~~~~~~~~~~~~~~~~~~~~~~~~~~~~~~~~~~~~~~
\par
We first assume the replica symmetry (RS). 
Using the replica symmetric partition function $Z_{RS}(\beta)$, we find the replica symmetric free energy as 
\begin{eqnarray}
  f_{RS}(\beta)
  &=& -\frac 1{\beta nM} \ln \mathbb{E}_{\mathcal{A},\vec{J}}[Z_{RS}(\beta)^n] 
      \label{eq:replica_trick} \\
  &=& \frac 12 - \frac 1{\beta} \mathop{\rm extr}_{\pi, \hat{\pi}} \biggl[ 
      \ln \cosh \frac{\beta}2 
      - \bar{K} \int_{-1}^1 dx \pi(x) \int_{-1}^1 d\hat{x} \hat{\pi}(\hat{x}) \ln ( 1 + x \hat{x} ) \nonumber \\
  & & + \sum_{K} P_K(K) \biggl( \prod_{k=1}^K \int_{-1}^1 dx_k \pi(x_k) \biggr) 
      \mathbb{E}_J \biggl[ \ln \biggl( 1 + \biggl( \tanh  \frac{\beta J}2 \biggr) \prod_{k=1}^K x_k \biggr) \biggr] \nonumber \\
  & & + \frac{\bar{K}}{\bar{C}} \sum_{C} P_C(C) \biggl( \prod_{c=1}^C \int_{-1}^1 d\hat{x}_c \hat{\pi}(\hat{x}_c) \biggr) 
      \ln \biggl( \sum_{\sigma=\pm 1} \prod_{c=1}^C [ 1 + \sigma \hat{x}_c] \biggr) 
      \biggr], 
      \label{eq:fRS}
\end{eqnarray}
where the parameters are determined by the saddle-point equations obtained by calculating functional variations: 
\begin{eqnarray}
  \pi(x) 
  &=& \sum_{C} \frac{C}{\bar{C}} P_C(C) 
      \biggl( \prod_{c=1}^{C-1} \int_{-1}^1 d\hat{x}_c \hat{\pi}(\hat{x}_c) \biggr) 
      \delta \biggl( x - \tanh \biggl[ \sum_{c=1}^{C-1} \tanh^{-1} \hat{x}_c \biggr] \biggr), 
      \label{eq:rs_sp_pi} \\
  \hat{\pi}(\hat{x}) 
  &=& \sum_{K} \frac{K}{\bar{K}} P_K(K) \biggl( \prod_{k=1}^{K-1} \int_{-1}^1 dx_k \pi(x_k) \biggr) 
      \mathbb{E}_J \biggl[ \delta \biggl( \hat{x} - \biggl( \tanh \frac{\beta J}2 \biggr) \prod_{k=1}^{K-1} x_k \biggr) \biggr], \nonumber \\
      \label{eq:rs_sp_pi^}
\end{eqnarray}
with $\bar{K} = \sum_{K} K P_K(K)$ and $\bar{C} = \sum_{C} C P_C(C)$ (See the outline of the derivation in \ref{app:f}). 
We can obtain the distortion, which is reproduction errors, $u_{RS}(\beta) = \partial [ \beta f_{RS}(\beta) ] / \partial \beta$ and 
the replica symmetric (RS) entropy $s_{RS}(\beta) = \beta [ u_{RS}(\beta) - f_{RS}(\beta) ]$. 
\par
For arbitrary $P_K(K)$, $P_C(C)$ and $\beta$, $\pi(x)=\delta(x)$ and $\hat{\pi}(\hat{x}) = \delta(\hat{x})$ are always solutions of the saddle-point equations (\ref{eq:rs_sp_pi}) and (\ref{eq:rs_sp_pi^}).
These are correspond the paramagnetic solution. 
The paramagnetic free-energy, internal energy and entropy are given by 
$f_{PARA}(\beta) = \frac 12 - \frac 1{\beta} \ln \cosh \frac{\beta}2 - \frac R{\beta} \ln 2$, 
$u_{PARA}(\beta) = \frac 12 - \frac12 \tanh \frac{\beta}2$ and 
$s_{PARA}(\beta) = \ln \cosh \frac {\beta}2 - \frac{\beta}2 \tanh \frac{\beta}2 + R \ln 2$, respectively. 
However, this RS solution takes negative entropy while $R \ln 2 < \frac \beta 2 \tanh \frac \beta 2 - \ln \cosh \frac \beta 2$. 
Especially, when the inverse temperature $\beta\to\infty$, the RS entropy becomes $s_{RS}(\beta)=(R-1)\ln 2$. 
This means we have to look for the true solution beyond the RS ansatz for $R \le 1$.

\subsection{One-step replica symmetry breaking solution}
%~~~~~~~~~~~~~~~~~~~~~~~~~~~~~~~~~~~~~~~~~~~~~~~~~~~~~~~~~~~~~~~~~~~~~
\par
The replica symmetric breaking (RSB) theory for sparse systems is still under development \cite{Wong1988, deDominicis1989, Lai1990, Goldschmidt, Monasson1998, Parisi2002}. 
Therefore, as a first approach we introduce the frozen RSB to produce a solution with non-negative entropy \cite{Murayama2003, Kabashima1999, Vicente2000}. 
The frozen RSB method is a limited version of full one-step RSB (1RSB) and includes the RS method as a special case. 
In this 1RSB scheme, $n$ replicas are divided into $n/m$ groups which contain $m$ replicas each. 
The symmetry breaking parameter $m$ was found to be $m=\beta_g/\beta$, 
where $\beta_g$ is a inverse temperature at which the replica symmetric entropy vanishes, i.e., $s_{RS}(\beta_g)=0$ (See \ref{app:1RSB}). 
This 1RSB scheme gives the exact solution for the random energy model (REM) \cite{Murayama2003, Derrida1981}. 
For $\beta>\beta_g$, the 1RSB free energy becomes $f_{1RSB}(\beta) = f_{RS}(\beta_g)$. 
It can be ragarded as a constant with respect to the inverse temperature $\beta$. 
We assume that the 1RSB scheme is enough good to approximate the solution even if $\bar{K}$ and $\bar{C}$ are finite. 
Under this assumption, the distortion $D$ is simply given by $D=\lim_{\beta\to\infty} u_{1RSB}(\beta) = u_{RS}(\beta_g)$.

\section{Results and discussion}
%~~~~~~~~~~~~~~~~~~~~~~~~~~~~~~~~~~~~~~~~~~~~~~~~~~~~~~~~~~~~~~~~~~~~~

\subsection{Basic results}
%~~~~~~~~~~~~~~~~~~~~~~~~~~~~~~~~~~~~~~~~~~~~~~~~~~~~~~~~~~~~~~~~~~~~~
\par
In large $\bar{K}$ and $\bar{C}$ limit, 
there are no other solutions except $\pi(x)=\delta(x)$ and $\hat{\pi}(\hat{x}) = \delta(\hat{x})$ for the saddle-point equations. 
We then found the relationship 
\begin{equation}
  R=1-h_2(D), 
  \label{eq:rdf}
\end{equation}
from 
$s_{RS}(\beta_g) = \ln \cosh \frac {\beta_g}2 - \frac{\beta_g}2 \tanh \frac{\beta_g}2 + R \ln 2 = 0$ and 
$D = u_{RS}(\beta_g) = \frac 12 -\frac 12 \tanh \frac {\beta_g}2$. 
\par
\begin{figure}[t]%[htbp]
  \vspace{5mm}
  \begin{center}
    \includegraphics[width=.40\linewidth,keepaspectratio]{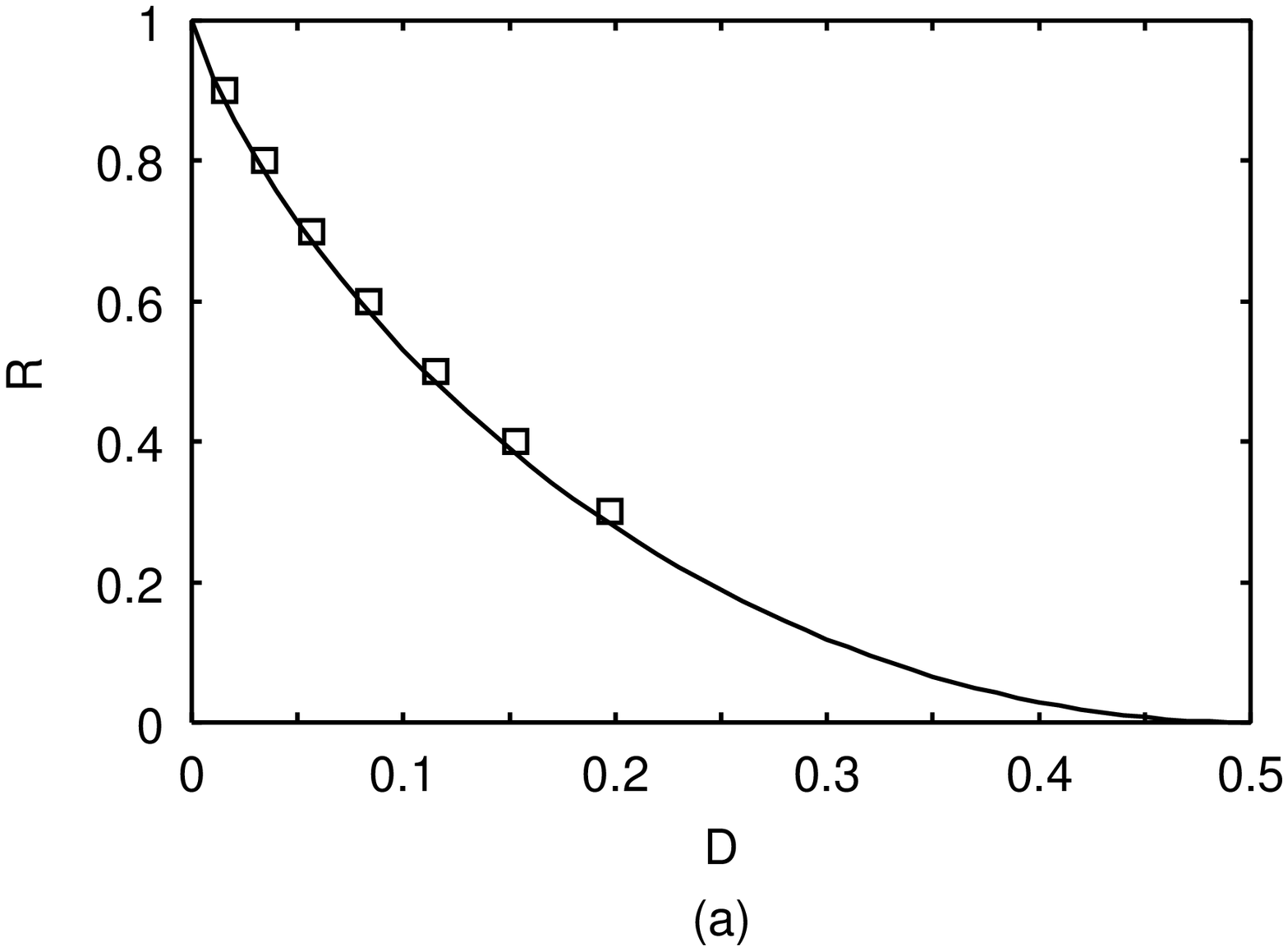} \hspace{0.05\linewidth}
    \includegraphics[width=.40\linewidth,keepaspectratio]{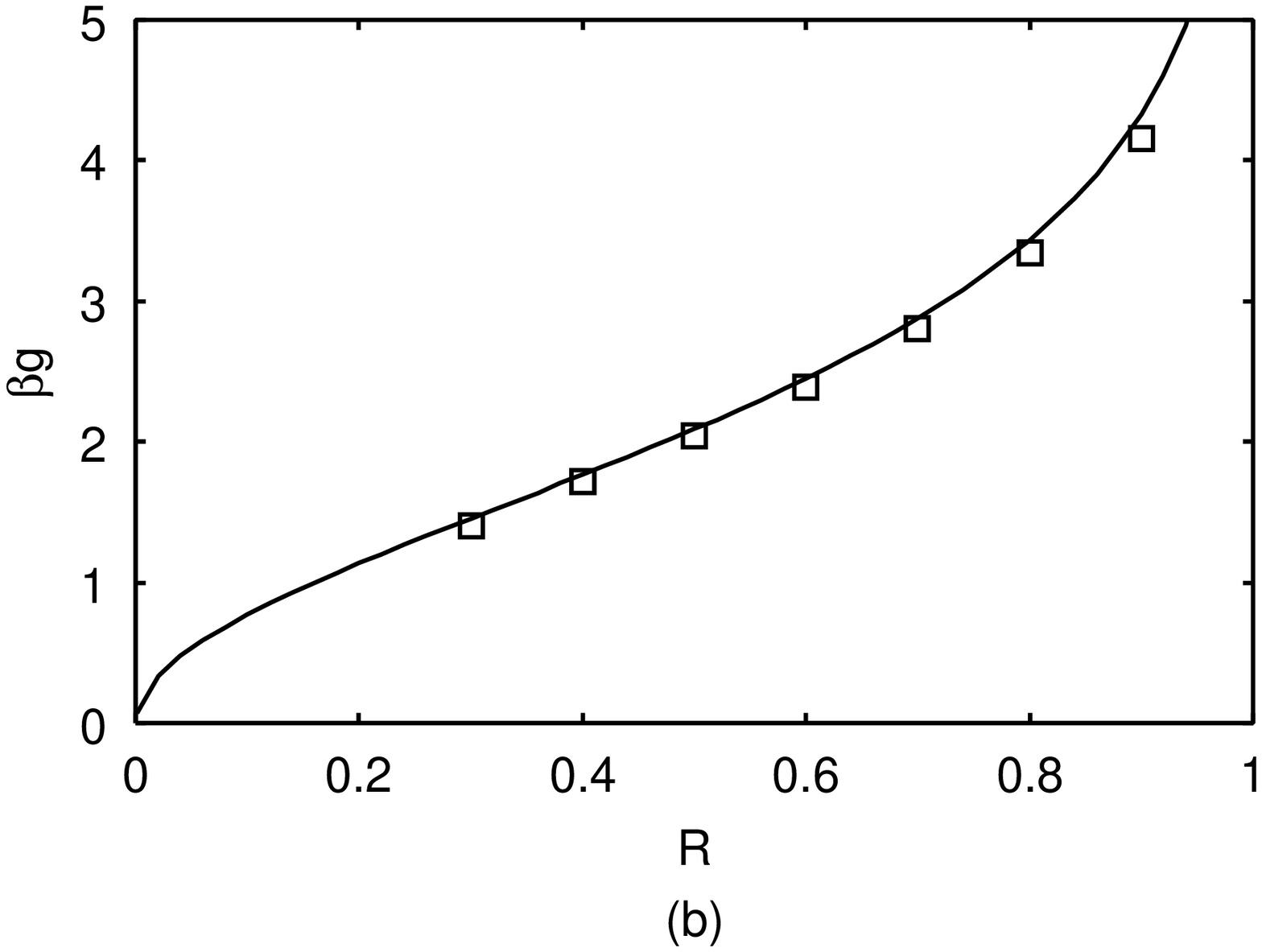} 
    \caption{Example of numerical solutions for finite connectivity systems with $P_K(K)=\delta_{K,2}$ and $P_C(C)=\tilde{\mathcal{P}}_{C}(C)$. 
             (a) Rate distortion performance for $r=0.3, 0.4, \cdots, 0.9$ (squares). 
                 The solid line denotes the rate distortion performance in large $\bar{K}$ and $\bar{C}$ limit, 
                 which coincides with the Shannon bound. 
             (b) Inverse temperature $\beta_g$ for for $r=0.3, 0.4, \cdots, 0.9$ (squares). 
                 The solid line denotes the inverse temperature $\beta_c$, which is defined by $s_{PARA}(\beta_c)=0$. 
    }
    \label{fig:rdp}
  \end{center}
\end{figure}
\par
In finite $\bar{K}$ and $\bar{C}$ case, 
the solutions $\pi(x)=\delta(x)$ and $\hat{\pi}(\hat{x}) = \delta(\hat{x})$ also exist, 
but these are no longer stable \cite{Murayama2003}. 
We have to solve the equations (\ref{eq:rs_sp_pi}) and (\ref{eq:rs_sp_pi^}) numerically. 
We choose the proper value of the inverse temperature $\beta_g$ which gives $s_{RS}(\beta_g)=0$ by using the numerical results of the saddle-point equations. 
Since the distortion $D$ can be evaluated from $D=u_{RS}(\beta_g)$, 
we can also obtain the relation between the code rate $R=\bar{K}/\bar{C}$ and the distortion $D$ in the finite connectivity systems. 
\par
As one of the simplest examples to treat the arbitrary code rate, 
we here introduce degree distributions 
$P_K(K) = \delta_{K,2}$ and 
$P_C(C) = \frac{7r-2}{5r} \delta_{C,2} + \frac{2(1-r)}{5r} \delta_{C,7} (\equiv \tilde{\mathcal{P}}_{C}(C))$, 
which are valid for $\frac 27 \le r \le 1$. 
Here, $\delta_{m,n}$ denotes Kronecker's delta taking 1 if $m=n$ and 0 otherwise. 
In this case, we can adjust the code rate $R(=r)$ via the parameter $r$. 
We apply the Monte-Carlo integration to solve the saddle-point equations. 
Figure \ref{fig:rdp} (a) shows the rate-distortion performance of this system. 
Figure \ref{fig:rdp} (b) shows the inverse temperature $\beta_g$, which is a root of the replica symmetric entropy $s_{RS}(\beta_g)=0$. 
\par
\begin{figure}[t]%[htbp]
  \vspace{5mm}
  \begin{center}
    \small
    \includegraphics[width=.35\linewidth,keepaspectratio]{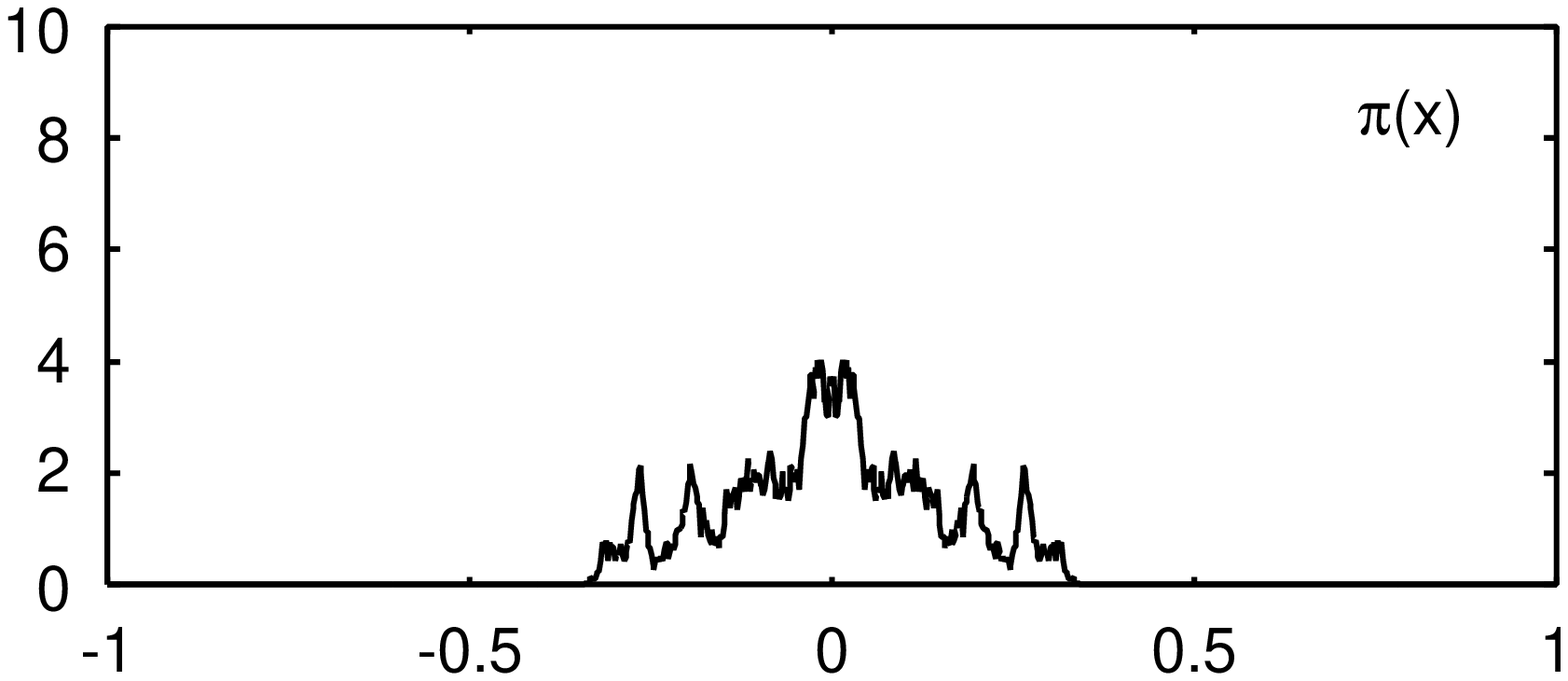} \hspace{0.05\linewidth}
    \includegraphics[width=.35\linewidth,keepaspectratio]{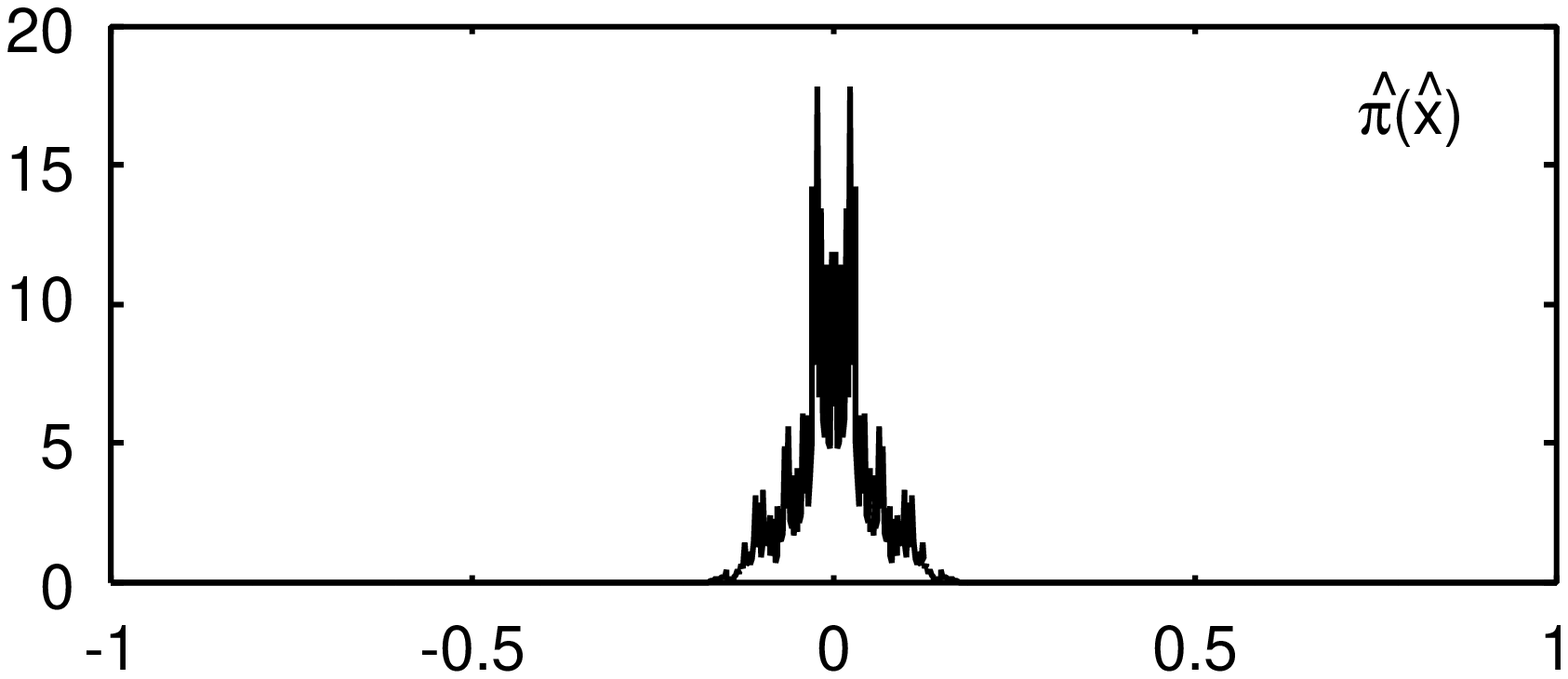} \\
    (a) $P_{K}(K)=\delta_{K,2}$, $P_{C}(C)=\delta_{C,4}$ \\ \vspace{2mm}
    \includegraphics[width=.35\linewidth,keepaspectratio]{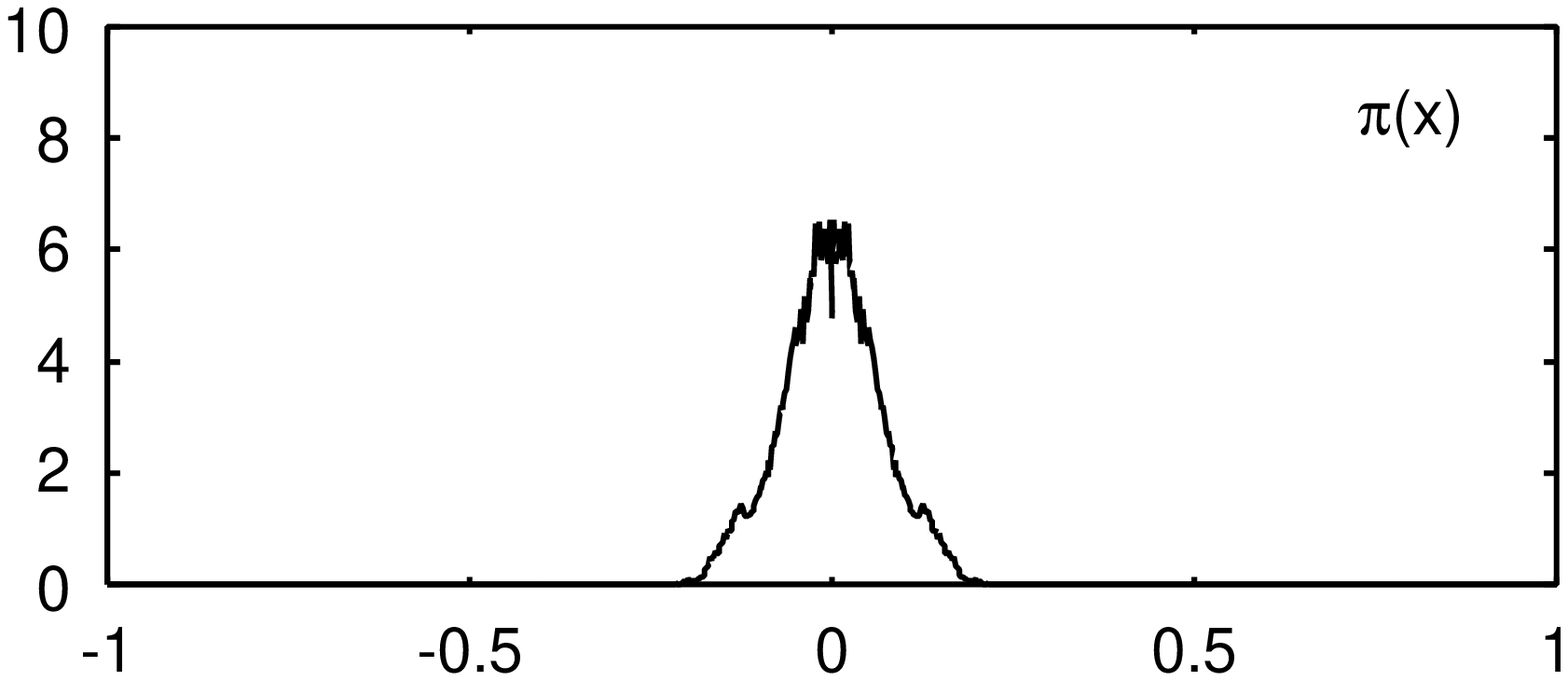} \hspace{0.05\linewidth}
    \includegraphics[width=.35\linewidth,keepaspectratio]{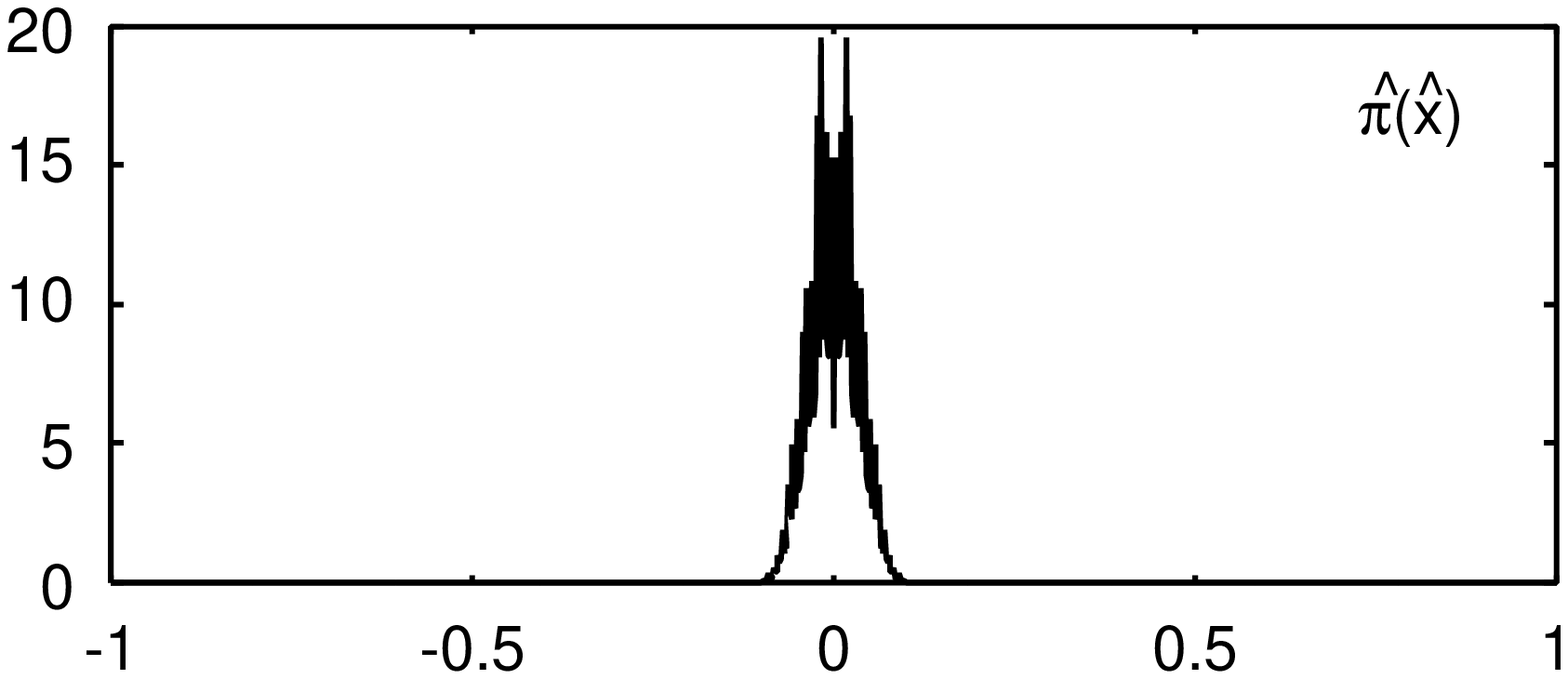} \\
    (b) $P_{K}(K)=\delta_{K,2}$, $P_{C}(C)=\mathcal{P}_{C}(C)$ \\ \vspace{2mm}
    \includegraphics[width=.35\linewidth,keepaspectratio]{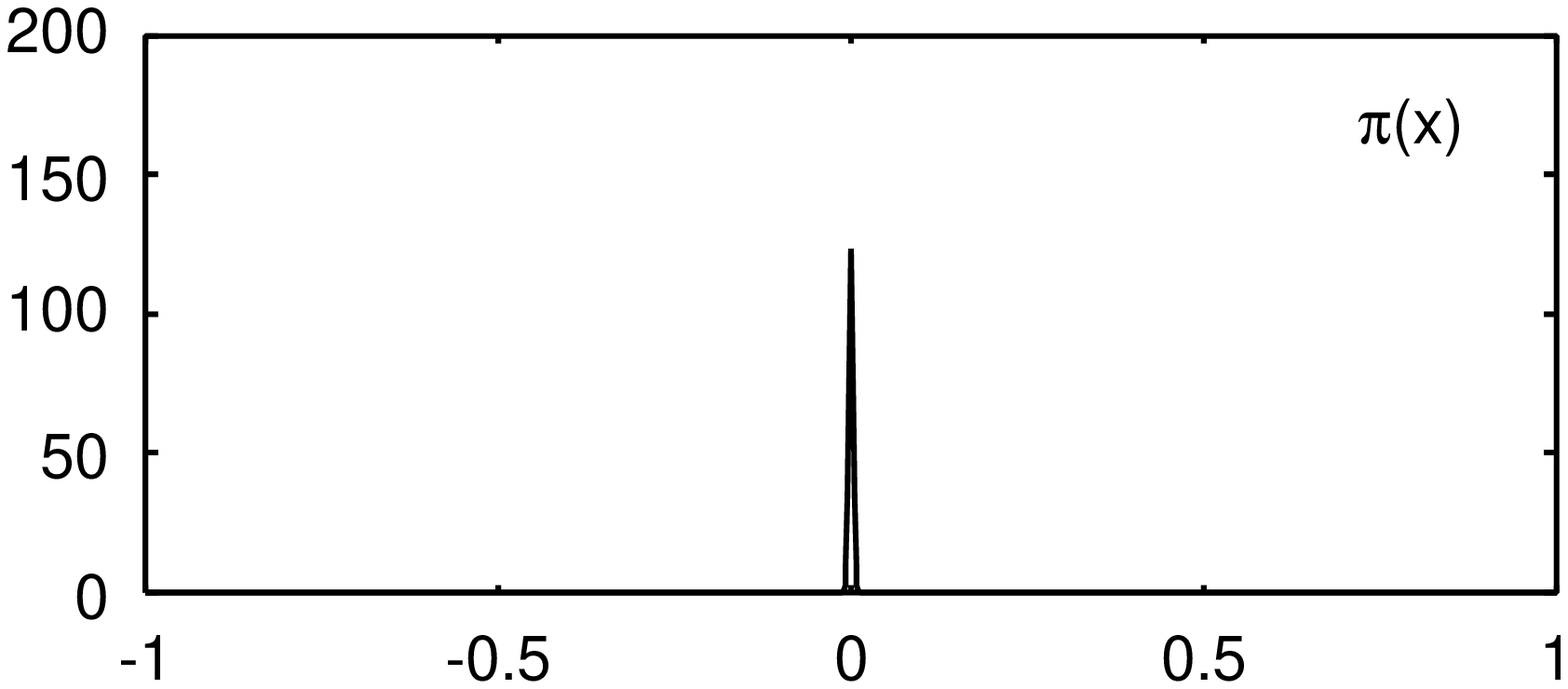} \hspace{0.05\linewidth}
    \includegraphics[width=.35\linewidth,keepaspectratio]{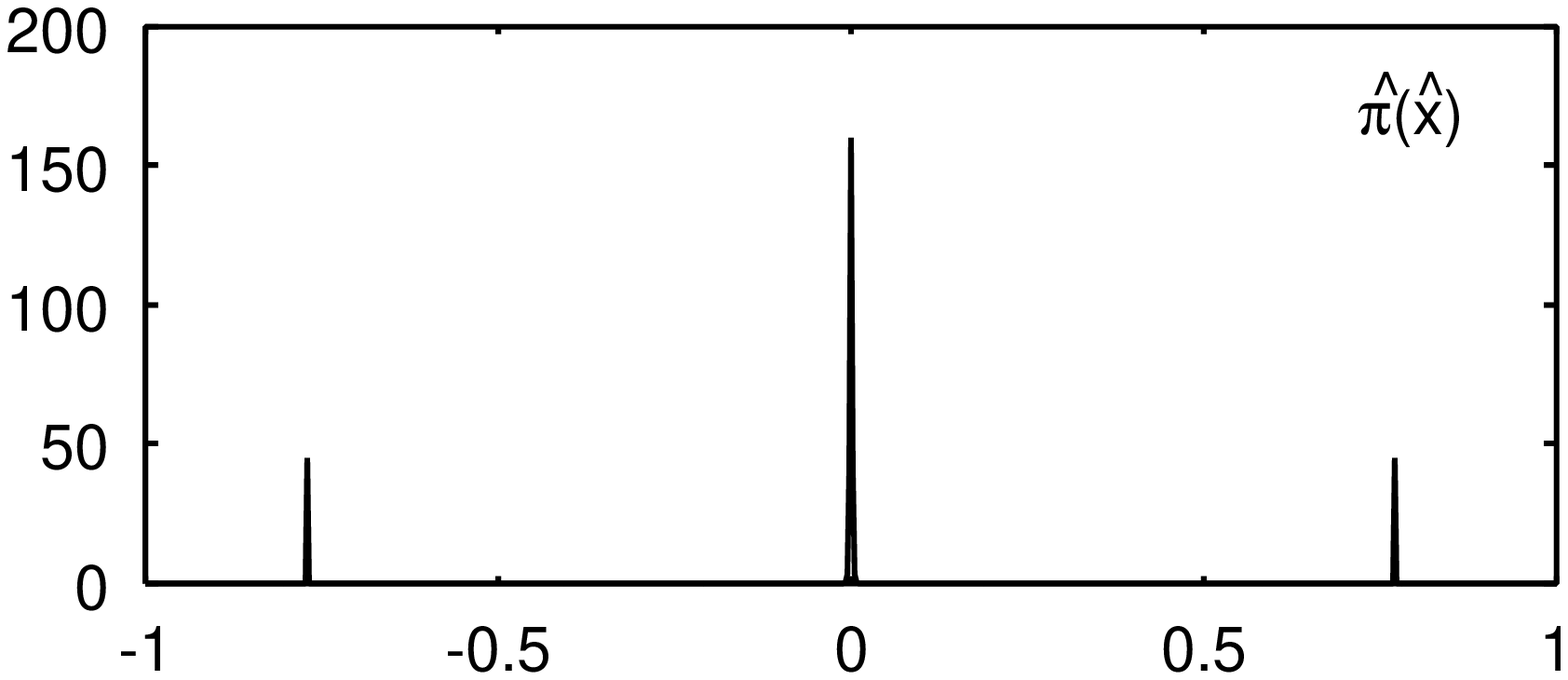} \\
    (c) $P_{K}(K)=\mathcal{P}_{K}(K)$, $P_{C}(C)=\delta_{C,4}$ 
    \normalsize
    \caption{Snapshots of the order functions $\pi(x)$ and $\hat{\pi}(\hat{x})$. 
             (b) a regular case, 
             (a) a check-regular and bit-irregular case, and 
             (c) a check-irregular and bit-regular case.}
    \label{fig:solution}
  \end{center}
  %P_K(K) : check degree distribution
  %P_C(C) : bit degree distribution
\end{figure}

\subsection{Some typical irregular constructions}
%~~~~~~~~~~~~~~~~~~~~~~~~~~~~~~~~~~~~~~~~~~~~~~~~~~~~~~~~~~~~~~~~~~~~~
\par
We next apply some degree distributions as typical examples. 
It should be noted that these distributions discussed here are not optimized but heuristically chosen. 
All three examples have the code rate $R=1/2$. 
\par
Firstly, we consider a regular case characterized as $P_K(K)=\delta_{K,2}$ and $P_C(C)=\delta_{C,4}$. 
Figure \ref{fig:solution} (a) shows stable solutions $\pi(x)$ and $\hat{\pi}(\hat{x})$ of the saddle-point equations for this case. 
It can be confirmed that the functions $\pi(x)$ and $\hat{\pi}(\hat{x})$ are broad in shape. 
In this case, the distortion becomes $D=0.116$. 
The Shannon bound is $D_{SB}=0.1100$. 
\par
Secondly, We treat a check-regular and bit-irregular case 
whose degree distributions are defined as $P_K(K)=\delta_{K,2}$ and $P_C(C)=\mathcal{P}_C(C)$, where 
\begin{eqnarray}
  \mathcal{P}_{C}(C)
  &=& 0.04 \delta_{C,1} + 0.15 \delta_{C,2} + 0.22 \delta_{C,3} + 0.22 \delta_{C,4} \nonumber \\
  & & + 0.18 \delta_{C,5} + 0.11 \delta_{C,6} + 0.08 \delta_{C,7}. 
\end{eqnarray}
This $\mathcal{P}_C(C)$ is a rough approximation of the Poissonian distribution $e^{-\lambda} \lambda^{C-1} / (C-1)!$ with $\lambda = 3$. 
The distortion is $D=0.115$ for this case. 
It represents an ensemble which have at least one non-zero element in each row. 
In the check-regular case, when we choose the non-zero elements randomly, 
there exists some columns whose elements are all zero. 
In such a situation, the code rate essentially becomes small. 
\par
Lastly, for a check-irregular and bit-regular case, 
we apply $P_K(K)=\mathcal{P}_K(K)$ and $P_C(C)=\delta_{C,4}$, where 
\begin{eqnarray}
  \mathcal{P}_{K}(K)
  = 0.36 \delta_{K,1} + 0.36 \delta_{K,2} + 0.20 \delta_{K,3} + 0.08 \delta_{K,4}. 
\end{eqnarray}
This $\mathcal{P}_K(K)$ is a rough approximation of the Poissonian distribution $e^{-\lambda} \lambda^{K-1} / (K-1)!$ with $\lambda = 1$. 
The reason why we consider this distribution is same to $\mathcal{P}_{C}(C)$. 
In this case, the distortion becomes $D=0.115$. 
These three kinds of distributions give almost same distortion. 
\par
Figure \ref{fig:solution} (b) and (c) show stable solutions for these irregular cases. 
It can be confirmed that the distribution $\pi(x)$ and $\hat{\pi}(\hat{x})$ become a little bit narrow than the regular case. 
It is considered that the distortion can become small due to this.

\section{Conclusions}
%~~~~~~~~~~~~~~~~~~~~~~~~~~~~~~~~~~~~~~~~~~~~~~~~~~~~~~~~~~~~~~~~~~~~~
\par
We evaluate typical performance of LDGM codes with irregular bit and check degree distributions by applying the replica method under 1RSB ansatz. 
Our result shows that we can use an arbitrary code rate. 
It might be possible to investigate suboptimal irregular degree distributions 
by using the hill-climbing approach similar to the case of the density evolution \cite{Richardson2001a, Richardson2001b}. 
\par
In the practical point of view, it must be important to evaluate some polynomial time encoding algorithms with arbitrary degree distributions. 
It should be noted that the analysis addressed here is based on an exact calculation of the encoder's definition. 
Therefore it can be considered that the distortion obtained by this analysis provides the theoretical limit for given check and bit degree distributions. 
\par
Recently, the cavity method was introduced to evaluate the typical performance \cite{Ciliberti2005}. 
Since the cavity method does not need the replica trick, it might be able to avoid some assumptions. 
Applying the cavity method to this problem is also important and is a part of our future work.

\ack
%~~~~~~~~~~~~~~~~~~~~~~~~~~~~~~~~~~~~~~~~~~~~~~~~~~~~~~~~~~~~~~~~~~~~~
The author would like to thank Kazutaka Nakamura, Tatsuto Murayama and Yoshiyuki Kabashima for their helpful comments. 
We also thank Toru Yano for giving us valuable preprints. 
This work was partially supported by 
a Grant-in-Aid for Encouragement of Young Scientists (B) No. 18700230 
from the Ministry of Education, Culture, Sports, Science and Technology of Japan.

\appendix
\section{Derivation of replica symmetric free energy \label{app:f}}
%~~~~~~~~~~~~~~~~~~~~~~~~~~~~~~~~~~~~~~~~~~~~~~~~~~~~~~~~~~~~~~~~~~~~~

We assume that the event $\mathcal{D}_{i_1, \cdots, i_{K_\mu}}^\mu = 1$ occurs independently for every row $\mu$. 
We then have 
\begin{eqnarray}
  & & \mathbb{P} (\mathcal{D}_{i_1, \cdots, i_{K_\mu}}^\mu = 1) = p_\mu, \\
  & & \mathbb{P} (\mathcal{D}_{i_1, \cdots, i_{K_\mu}}^\mu = 0) = 1 - p_\mu, 
\end{eqnarray}
where $\mathbb{P}(\cdots)$ denotes the probability of the event $(\cdots)$ and $p_\mu = ({}_{K_\mu}^N)^{-1} \simeq K_\mu!/N^{K_\mu}$. 
Introducing the constraint concerning the column (\ref{eq:column_constraint}) by using Dirac's delta function,
the ensemble average over the codes is represented as 
\begin{eqnarray}
  & & \mathbb{E}_{\mathcal{A}}[ (\cdots) ] \nonumber \\
  & & = \biggl( \sum_{\{K_\mu\}} \prod_{\mu=1}^M P_K(K_\mu) \biggr) 
      \biggl( \sum_{\{C_i\}}\prod_{i=1}^N P_C(C_i) \biggr) \nonumber \\
  & & \;\;\;\; \times \frac 1{\mathcal{N_D}} \mathbb{E}_{\mathcal{D}} \biggl[ 
      \biggl\{ \prod_{i=1}^N \delta\biggl(  \sum_{\mu=1}^M \sum_{\langle i_1=i, i_2, \cdots, i_{K_\mu} \rangle} \!\!\!\! \mathcal{D}_{i_1=i, i_2, \cdots, i_{K_\mu}}^\mu ; C_i \biggr) \biggr\} 
      (\cdots) \biggr], \nonumber \\
  & & =\biggl( \sum_{\{K_\mu\}} \prod_{\mu=1}^M P_K(K_\mu) \biggr) 
      \biggl( \sum_{\{C_i\}}\prod_{i=1}^N P_C(C_i) \biggr) \nonumber \\
  & & \;\;\;\; \times \frac 1{\mathcal{N_D}} \mathbb{E}_{\mathcal{D}} \biggl[ 
      \biggl\{ \prod_{i=1}^N 
      \oint \frac{dZ_i}{2\pi i}\frac1{Z_i^{C_i+1}} \prod_{\mu=1}^N \prod_{\langle i_1=i, i_2, \cdots, i_{K_\mu} \rangle} \!\!\!\! Z_i^{\mathcal{D}_{i_1=i, i_2, \cdots, i_{K_\mu}}^\mu}
      \biggr\} 
      (\cdots) \biggr], 
\end{eqnarray}
where $\mathbb{E}_{\mathcal{D}}$ denotes the average over the connectivity parameter. 
Observing that $\sum_{\langle i_1, \cdots, i_{K_\mu} \rangle} (\cdots) = \frac 1{K_\mu!}(\sum_i (\cdots) )^{K_\mu}$ for large $N$, 
the normalization constant $\mathcal{N_D}$ is given by 
\begin{eqnarray}
  \mathcal{N_D}
  = \mathbb{E}_{\mathcal{D}} \biggl[ \prod_{i=1}^N \delta\biggl(  \sum_{\mu=1}^M \sum_{\langle i_1=i, i_2, \cdots, i_{K_\mu} \rangle} \!\!\!\! \mathcal{D}_{i_1=i, i_2, \cdots, i_{K_\mu}}^\mu ; C_i \biggr) \biggr] 
  = \frac{(N\bar{C})!}{\displaystyle N^{N\bar{C}} \prod_{i=1}^N C_i!}. 
\end{eqnarray}
To evaluate the free energy, we calculate the replicated partition function: 
\begin{eqnarray}
  & & \mathbb{E}_{\mathcal{A},\vec{J}}[Z(\beta)^n] \nonumber \\
  & & = e^{-\frac{nM\beta}2} \mathbb{E}_{\mathcal{A},\vec{J}} \Biggl[ \sum_{\vec{s}^1,\cdots,\vec{s}^n} \exp \biggl[ 
      \frac \beta 2 \sum_{\mu=1}^M \sum_{\langle i_1, \cdots, i_{K_\mu} \rangle} \!\!\!\! \mathcal{D}_{i_1, \cdots, i_{K_\mu}}^\mu J_\mu \sum_{\alpha=1}^n s_{i_1}^\alpha \cdots s_{i_{K_\mu}}^\alpha \biggr\} 
      \biggr] \Biggr] \nonumber \\
  & & = e^{-\frac{nM\beta}2} 
      \biggl( \sum_{\{K_\mu\}} \prod_{\mu=1}^M P_K(K_\mu) \biggr) 
      \biggl( \sum_{\{C_i\}} \prod_{i=1}^N P_C(C_i) \biggr) \nonumber \\
  & & \;\;\;\; 
      \times \frac 1{\mathcal{N_D}} \biggl( \prod_{i=1}^N \oint \frac{dZ_i}{2\pi i}\frac1{Z_i^{C_i+1}} \biggr) 
      \prod_{\mu=1}^M \biggl( p_\mu \sum_{\langle i_1, \cdots, i_{K_\mu} \rangle} (\cosh \frac{\beta}2)^n Z_{i_1} \cdots Z_{i_{K_\mu}} \nonumber \\
  & & \;\;\;\; \qquad 
      + p_\mu \sum_{m=1}^n \sum_{\langle \alpha_1, \cdots, \alpha_m \rangle} (\cosh \frac{\beta}2)^n \mathbb{E}_J \biggl[ (\tanh \frac{\beta J}2)^m \biggr] \nonumber \\
  & & \;\;\;\; \qquad 
      \times \sum_{\langle i_1, \cdots, i_{K_\mu} \rangle} (Z_{i_1} s_{i_1}^{\alpha_1} \cdots s_{i_1}^{\alpha_m}) \cdots (Z_{i_{K_\mu}} s_{i_{K_\mu}}^{\alpha_1} \cdots s_{i_{K_\mu}}^{\alpha_m}) 
      \biggr). 
\end{eqnarray}
We next introduce order parameters $q_{\alpha_1,\cdots,\alpha_m}$ and $q_0$, defined by 
\begin{eqnarray}
  & & q_{\alpha_1,\cdots,\alpha_m} = \frac 1N \sum_{i=1}^N Z_i s_i^{\alpha_1} \cdots s_i^{\alpha_m}, \\
  & & q_0 = \frac 1N \sum_{i=1}^N Z_i. 
\end{eqnarray}
Using the Fourier expression of the Dirac delta function, we find 
\begin{eqnarray}
  & & \mathbb{E}_{\mathcal{A},\vec{J}}[Z(\beta)^n] \nonumber \\
  & & = e^{-\frac{nM\beta}2}
      \biggl( \int \frac{dq_0 d\hat{q}_0}{2\pi} \biggr) 
      \biggl( \prod_{\langle \alpha_1 \rangle} \int \frac{dq_{\alpha_1} d\hat{q}_{\alpha_1}}{2\pi} \biggr) \cdots 
      \biggl( \!\!\! \prod_{\langle \alpha_1, \cdots, \alpha_n \rangle} \int \frac{dq_{\alpha_1, \cdots ,\alpha_n} d\hat{q}_{\alpha_1, \cdots ,\alpha_n}}{2\pi} \biggr) \nonumber \\
  & & \;\;\;\; 
      \times 
      \biggl( \sum_{\{K_\mu\}} \prod_{\mu=1}^M P_K(K_\mu) \biggr) 
      \biggl( \sum_{\{C_i\}} \prod_{i=1}^N P_C(C_i) \biggr) 
      \frac 1{\mathcal{N_D}} \biggl( \prod_{i=1}^N \oint \frac{dZ_i}{2\pi i}\frac1{Z_i^{C_i+1}} \biggr) 
      \nonumber \\
  & & \;\;\;\; 
      \times \exp \biggl[
      -N \biggl\{ q_0 \hat{q}_0 + \cdots + \sum_{\langle \alpha_1, \cdots, \alpha_n \rangle} q_{\alpha_1, \cdots ,\alpha_n} \hat{q}_{\alpha_1, \cdots ,\alpha_n} \biggr\}
      \nonumber \\
  & & \;\;\;\; \qquad 
      + \hat{q}_0 \sum_{i=1}^N Z_i + \cdots + \sum_{\langle \alpha_1, \cdots, \alpha_n \rangle} \hat{q}_{\alpha_1, \cdots ,\alpha_n} \sum_{i=1}^N Z_i s_i^{\alpha_1} \cdots s_i^{\alpha_n} \biggr] 
      \nonumber \\
  & & \;\;\;\; 
      \times \prod_{\mu=1}^M \biggl( T_0 q_0^{K_\mu} + \sum_{m=1}^n \sum_{\langle \alpha_1, \cdots, \alpha_m \rangle} T_m (q_{\alpha_1, \cdots ,\alpha_m})^{K_\mu} \biggr) ,
\end{eqnarray}
with $T_m=(\cosh \frac{\beta}2)^n \mathbb{E}_J [ (\tanh \frac{\beta J}2)^m ]$. 
To proceed further, we introduce the replica-symmetric (RS) assumption: 
\begin{eqnarray}
  && q_{\alpha_1,\cdots,\alpha_m} = q \int_{-1}^1 dx \pi(x) x^m, \\
  && \hat{q}_{\alpha_1,\cdots,\alpha_m} = \hat{q} \int_{-1}^1 d\hat{x} \hat{\pi}(\hat{x}) \hat{x}^m, 
\end{eqnarray}
where $\pi(x) \ge 0$, $\hat{\pi}(\hat{x}) \ge 0$ and $\int_{-1}^1 dx \pi(x) = \int_{-1}^1 d\hat{x} \hat{\pi}(\hat{x}) =1$. 
This assumption means that the order parameters depend only on the number of indices. 
We write the replica symmetric partition function as $Z_{RS}(\beta)$. 
Using the integral form of the Dirac's delta function, we obtain 
\begin{eqnarray}
  & & \mathbb{E}_{\mathcal{A},\vec{J}}[Z_{RS}(\beta)^n] \nonumber \\
  & & =
      \mathop{\rm extr}_{\pi, \hat{\pi}, q, \hat{q}} 
      \frac {e^{-\frac{nM\beta}2}}{\mathcal{N_D}}
      \biggl( \sum_{\{K_\mu\}} \prod_{\mu=1}^M P_K(K_\mu) \biggr) 
      \biggl( \sum_{\{C_i\}} \prod_{i=1}^N P_C(C_i) \biggr) \nonumber \\
  & & \;\;\;\; 
      \times 
      \biggl( \prod_{i=1}^N \biggl\{ \frac{\hat{q}^{C_i}}{C_i!} 
      \biggl( \prod_{c=1}^{C_i} \int_{-1}^1 d\hat{x}_c \hat{\pi}(\hat{x}_c) \biggr)
      \biggl( \sum_{\sigma = \pm 1} \prod_{c=1}^{C_i} (1+\sigma \hat{x}_c) \biggr)^n \biggr\} \biggr) \nonumber \\
  & & \;\;\;\; 
      \times 
      \exp \biggl[ -N q \hat{q} \int_{-1}^1 dx \pi(x) \int_{-1}^1 d\hat{x} \hat{\pi}(\hat{x}) (1+x\hat{x})^n \biggr] \nonumber \\
  & & \;\;\;\; 
      \times \prod_{\mu=1}^M \biggl( q^{K_\mu} (\cosh \frac {\beta}2)^n \biggl( \prod_{k=1}^{K_\mu} \int_{-1}^1 dx_k \pi(x_k) \biggr) 
      \mathbb{E}_{J} \biggl[ \biggl( 1 + (\tanh \frac{\beta J}2) \prod_{k=1}^{K_\mu} x_k \biggr)^n \biggr] . \nonumber \\
  \label{eq:ZRS}
\end{eqnarray}
Finally, substituting this into (\ref{eq:replica_trick}) and taking the limit $n \to 0$, we arrive at (\ref{eq:fRS}). 
The saddle-point equations (\ref{eq:rs_sp_pi}) and (\ref{eq:rs_sp_pi^}) are simply obtained as the extremization condition of (\ref{eq:fRS}).

\section{One-step replica symmetry breaking solution \label{app:1RSB}}
%~~~~~~~~~~~~~~~~~~~~~~~~~~~~~~~~~~~~~~~~~~~~~~~~~~~~~~~~~~~~~~~~~~~~~

We follow the calculation of the reference \cite{Vicente2000}. 
We assume that the space of configuration is divided in $n/m$ groups with $m$ identical configurations in each. 
\begin{equation}
  \frac 1N \vec{s}^\alpha \cdot \vec{s}^\beta =
  \left\{
  \begin{array}{cl}
    1, & {\rm if} \; \alpha \; {\rm and} \; \beta \; {\rm are \; in \; the \; same \; group} \\
    q, & {\rm otherwise} \\
  \end{array}
  \right. .
  \label{eq:eb}
\end{equation}
Using this ergodicity breaking assumption, the 1RSB replicated partition function becomes 
\begin{eqnarray}
  \mathbb{E}_{\mathcal{A},\vec{J}}[Z_{1RSB}(\beta)^n] |_{(\ref{eq:eb})} 
  &=& \mathbb{E}_{\mathcal{A},\vec{J}} \Biggl. \Biggl[ \Biggl( \sum_{\vec{s}} e^{-\beta \mathcal{H}(\vec{s},\vec{J})} \Biggr)^n \Biggr] \Biggr|_{(\ref{eq:eb})} \nonumber \\
  &=& \mathbb{E}_{\mathcal{A},\vec{J}} \Biggl[ \Biggl( \sum_{\vec{s}} e^{-\beta m \mathcal{H}(\vec{s},\vec{J})} \Biggr)^{n/m} \Biggr] \nonumber \\
  &=& \mathbb{E}_{\mathcal{A},\vec{J}}[Z_{RS}(\beta m)^{n/m}]. 
\end{eqnarray}
Then we obtain the 1RSB free energy as 
\begin{eqnarray}
  f_{1RSB}(\beta) 
  &=& - \frac 1{\beta} \mathbb{E}_{\mathcal{A},\vec{J}}[\ln Z_{1RSB}(\beta)] \nonumber \\
  &=& - \frac 1{\beta} \biggl( \frac{\partial}{\partial n} \mathbb{E}_{\mathcal{A},\vec{J}}[Z_{1RSB}(\beta)^n] \biggr) \biggl. \biggr|_{n=0} \nonumber \\
  &=& - \frac 1{\beta m} \biggl( \frac{\partial}{\partial (n/m)} \mathbb{E}_{\mathcal{A},\vec{J}}[Z_{RS}(\beta m)^{n/m}] \biggr) \biggl. \biggr|_{n/m=0} \nonumber \\
  &=& - \frac 1{\beta m} \mathbb{E}_{\mathcal{A},\vec{J}}[\ln Z_{RS}(\beta m)] \nonumber \\
  &=& f_{RS}(\beta m)
\end{eqnarray}
The symmetry breaking parameter $m$ should be determined to extremize the 1RSB free energy as 
\begin{eqnarray}
  \frac{\partial}{\partial m} f_{1RSB}(\beta) = 0. 
  \label{eq:condition_of_m}
\end{eqnarray}
The left hand side of this condition becomes 
\begin{eqnarray}
  \frac{\partial}{\partial m} f_{1RSB} (\beta) 
  &=& - \frac{\partial}{\partial m} \frac 1{\beta m} \mathbb{E}_{\mathcal{A},\vec{J}}[\ln Z_{RS}(\beta m)] \nonumber \\
  &=& - \frac 1m \biggl( \frac{\partial [ \mathbb{E}_{\mathcal{A},\vec{J}}[\ln Z_{RS}(\beta m)]] }{\partial (\beta m)} 
        - \frac 1{\beta m} \mathbb{E}_{\mathcal{A},\vec{J}}[\ln Z_{RS}(\beta m)] \biggr) \nonumber \\
  &=& \frac 1m \biggl( \frac{\partial [ (\beta m) f_{RS}(\beta m)] }{\partial (\beta m)} - f_{RS}(\beta m) \biggr) \nonumber \\
  &=& \frac 1{\beta m^2} s_{RS}(\beta m). 
\end{eqnarray}
Namely, the condition (\ref{eq:condition_of_m}) is equivalent to $s_{RS}(\beta m)=0$. 
Therefore, the symmetry breaking parameter is given by $m=\beta_g/\beta$ with $s_{RS}(\beta_g)=0$.

\section*{References}
%~~~~~~~~~~~~~~~~~~~~~~~~~~~~~~~~~~~~~~~~~~~~~~~~~~~~~~~~~~~~~~~~~~~~~

\end{document}